\documentclass[submitting]{nst}

\usepackage{subfigure,dcolumn}
\usepackage[T2A,T1]{fontenc}
\usepackage[russian,english]{babel}

\begin{document}

\preprint{APS/123-QED}

\title{Determination of neutron-skin thickness using configurational information entropy}\thanks{This work was supported by the National Natural Science Foundation of China (Nos. 11975091 and U1732135) and the Program for Innovative Research Team (in Science and Technology) in University of Henan Province, China (No. 21IRTSTHN011).}
\author[]{Chun-Wang Ma}
\email[Corresponding author, ]{machunwang@126.com}
\affiliation{School of Physics, Henan Normal University, Xinxiang 453007, China}
\affiliation{Key Laboratory Optoelectronic Sensing Integrated Application of Henan Province, Henan Normal University, Xinxiang 453007, China}

\author{Yi-Pu Liu}%
\affiliation{School of Physics, Henan Normal University, Xinxiang 453007, China}
\author{Hui-Ling Wei}%
\affiliation{School of Physics, Henan Normal University, Xinxiang 453007, China}
\author[]{Jie Pu}%
\affiliation{School of Physics, Henan Normal University, Xinxiang 453007, China}
\author{Kai-Xuan Cheng}%
\affiliation{School of Physics, Henan Normal University, Xinxiang 453007, China}
\author{Yu-Ting Wang}%
\affiliation{School of Physics, Henan Normal University, Xinxiang 453007, China}

\begin{abstract}
Configurational information entropy (CIE) theory was employed to determine the neutron skin thickness of neutron-rich calcium isotopes. The nuclear density distributions and fragment cross-sections in 350 MeV/u $^{40-60}$Ca + $^{9}$Be projectile fragmentation reactions were calculated using a modified statistical abrasion-ablation model. CIE quantities were determined from the nuclear density, isotopic, mass, and charge distributions. The linear correlations between the CIE determined using the isotopic, mass, and charge distributions and the neutron skin thickness of the projectile nucleus show that CIE provides new methods to extract the neutron skin thickness of neutron-rich nuclei.
\end{abstract}

\maketitle

\section{introduction}

Next generation radioactive nuclear beam facilities will provide new opportunities to explore extreme nuclei near and beyond drip lines. Nuclei with a large neutron excess can form exotic neutron skin or halo structures, which have attracted significant interest experimentally and theoretically for the past 30 years. The neutron skin thickness is defined as $\delta_\text{np} = \delta_\text{n} - \delta_\text{p}$, which denotes the difference between the point neutron and point proton root-mean-square (RMS) radii of a nucleus. Many methods have been developed to experimentally determine the neutron skin thickness. However, most of the models are indirect measurements and model dependent. Typical methods used to determine neutron skin thickness include the reaction cross section ($\sigma_\text{R}$), charge-changing cross section ($\sigma_\text{cc}$) \cite{CCCS14PRC,CCCS19PLB}, electric dipole polarizability \cite{Edipole15PRC}, photon multiplicity \cite{PhotonSkin15PRC}, the $\pi^{-}/\pi^{+}$ ratio or $\Sigma^-/\Sigma^+$ \cite{BALipi02,FZQ21CPL}, $^{3}$H/$^{3}$He ratio \cite{Yan19NST}, and $\alpha$ decay half-life time \cite{DecAlp17NST}. The projectile fragmentation reaction, which is the main experimental approach for studying rare isotopes, is suitable for determining the neutron skin thickness owing to the obvious experimental phenomena induced by the neutron skin structure \cite{HY20NST,JL21NST}. For example, isospin effects in the isotopic cross section\cite{Skinisospin}, neutron-abrasion cross section ($\sigma_\text{nabr}$) \cite{Ma08nabr}, neutron removal cross section \cite{PRL17nremval}, mirror nuclei ratio or isobaric ratio \cite{IYRm13PRC}, and isoscaling parameter ($\alpha$) \cite{SkinIsos15PRC}. Parity-violating electron scattering (PVS) is the only method used to determine neutron skin thickness that is model-independent. Ref. \cite{PVS_Ren13} reports a theoretical investigate of PVS for $^{48}$Ca and $^{208}$Pb, and a theoretical investigation of Bayesian approach is given in Ref.~\cite{XuJ21CPL}. Determining the neutron skin thickness of $^{48}$Ca and $^{208}$Pb is presently of significant interest and is listed in the U.S. 2015 Long Range Plan for Nuclear Science \cite{LRP2015USA}. The lead radius experiment (PREX) has been previously used to determine the neutron skin thickness of $^{208}$Pb \cite{Pb208PVS}. A recent PREX result indicates a much thicker neutron skin than previously predicted \cite{PREX2Pb}. Determining the neutron skin thickness of nuclei near the neutron drip line remains an important research outcome and is one of the most interesting topics in the new era of radioactive beam facilities.

Information entropy theory was established by C.E. Shannon \cite{Shannon}. The theory makes it possible to transform variables in a system into an exact information quantity \cite{CInfo16PLB} and has been used in various applications \cite{PPNP18Ma,PR-Chen18}. The first application of information entropy theory in heavy-ion reactions can be traced to the study of nuclear liquid-gas transition in nuclear multifragmentation \cite{YGMa99PRL}. Recent studies have extended it to study the information entropy carried by a single fragment produced in projectile fragmentation reactions, and revealed the scaling phenomenon of fragments covering a wide range of neutron excess  \cite{info15PLB,info16JPG,PPNP21Ma}. Configurational information entropy (CIE) was developed to quantify the information entropy of a physical distribution \cite{CEDef12PLB}, which connects the dynamical and informational contents of a physical system with localized configurations. Many applications of CIE methods can be found in Korteweg-de Vries (KdV) solitions, compact astrophysical systems, and scalar glueballs (see a brief introduction in Ref. \cite{PPNP18Ma}), theoretical research of new Higgs boson decay channels \cite{HiggsNewbranch20NPB}, deploying heavier eta meson states in AdS/QCD \cite{etaAds2021}, confinement/deconfinement transition in QCD \cite{QCDconftrans21}, quarkonium in a finite density plasma \cite{quakniumPlasma}, time evolution in physical systems \cite{timeRHIC2020,entropy20cpl}, etc. In projectile fragmentation reactions, fragment distributions show a sensitive dependence on the change in neutron density~\cite{IYRCaSkin14,Ma13finite,Ma13finite1}, which makes it possible to determine the neutron skin thickness of neutron-rich nuclei. In this study, the CIE method was adopted to quantify the CIE of nuclear density and fragment distributions in projectile fragmentation reactions. The analyzed data were generated using a modified statistical abrasion ablation (SAA) model, which is known to be a good model for describing the fragment cross sections of projectile fragmentation reactions \cite{FANG00,Ma09PRC}.

\section{Theories}
\label{Th:Sec}

\subsection{Modified Statistical Abrasion-Ablation Model}
\label{Th:SAA}
The modified statistical abrasion-ablation (SAA) model \cite{FANG00,Ma09PRC} can be used for projectile fragmentation reactions at both intermediate and high energies, which improves the original SAA model by Brohm and Schmidt \cite{Brohm94}. In quasi-free nucleon-nucleon collisions, the reaction is described as a two-step process: In the initial stage, the nucleons are described by a Glauber-type model as ``participants'' and ``spectators'', where the participants interact strongly in an overlapping region between the projectile and target, while the spectators move virtually undisturbed \cite{EIS54}. In the second stage, the excitation energy is compared to the separation energies of protons, neutrons, and $\alpha$ to determine the type of particle the prefragment can emit according to min($s_\text{p}, s_\text{n},s_{\alpha}$). After the de-excitation calculation, the cross sections of final fragments that are comparable to the measured fragments were obtained. The description of the modified SAA model is presented in Refs. \cite{PPNP18Ma,FANG00,Ma09PRC}.
The colliding nuclei are composed of many parallel tubes oriented along the beam direction. Their transverse motion is neglected, and the interactions between the tube pairs are independent. For a specific pair of interacting tubes, the absorption of the projectile neutrons and protons is assumed to be in a binomial distribution. At a given impact parameter $\mathbf{b}$, the transmission probabilities of neutrons (protons) of an infinitesimal tube in the projectile are calculated using
\begin{equation}
t_i(\mathbf{s}-\mathbf{b})=\exp\{-[D{_\text{n}^\text{T}}(\mathbf{s}-\mathbf{b})\sigma_{\text{n}i}+D{_\text{p}^\text{T}}(\mathbf{s}-\mathbf{b})\sigma_{\text{p}i}]\},
\end{equation}
where $D^\text{T}$ is the normalized integrated nuclear density distribution of the target along the beam direction for protons $\int \mathrm{d}^{2}sD_\text{p}^\text{T}=Z^\text{T}$ and neutrons $\int
\mathrm{d}^{2}sD_\text{n}^\text{T}=N^\text{T}$ ($N^\text{T}$ and $Z^\text{T}$ are the neutron and proton numbers of the target, respectively). $\mathbf{s}$ and $\mathbf{b}$ are defined in a plane perpendicular to the beam direction, and $\sigma_{i'i}$ denotes the free-space nucleon-nucleon cross sections ($i',i=\text{n}$ for neutrons and $i',i=\text{p}$ for protons) \cite{Cai98}. The average absorbed mass in the infinitesimal tube limit at a given $\mathit{\textbf{b}}$ is
\begin{eqnarray}
\langle\Delta A(b)\rangle=\int \mathrm{d}^{2}sD_\text{n}^\text{T}(\mathbf{s})[1-t_\text{n}(\mathbf{s}-\mathbf{b})] \nonumber\\
+\int \mathrm{d}^{2}sD_\text{p}^\text{P}(\mathbf{s})[1-t_\text{p}(\mathbf{s}-\mathbf{b})].
\end{eqnarray}
For a specific fragment, the production cross section can be calculated using
\begin{equation}
\sigma(\Delta N, \Delta Z)=\int \mathrm{d}^2b P(\Delta N, b)P(\Delta Z,b),
\end{equation}
where $P(\Delta N,b)$ and $P(\Delta Z, b)$ are the probability distributions of the abraded neutrons and protons at a given impact parameter $\mathit{b}$, respectively. $\sigma(\Delta N, \Delta Z)$ is the residual fragment after the abrasion stage, which is called the prefragment. The excitation energy of the prefragment is calculated as $E^*=13.3\langle\Delta A(b)\rangle$ MeV, where $\langle\Delta A(b)\rangle$ is the number of abraded nucleons from the projectile, and 13.3 MeV is the mean excitation energy owing to an abraded nucleon \cite{GAIM91}. In the second stage, the excitation energy is compared to the separation energies of protons, neutrons, and $\alpha$ to determine the type of particle the prefragment can emit according to min($s_\text{p}, s_\text{n},s_{\alpha}$). After the de-excitation calculation, the cross sections of final fragments that were comparable to the measured fragments were obtained.

Fermi-type density distributions were adopted for protons and neutrons in a nucleus, as shown in the equation below:

\begin{equation}
\rho_i(r)=\frac{\rho_i^0}{1+\mbox{exp}(\frac{r-C_i}{t_i/4.4})},~~~i=\text{n,\,p}
\end{equation}
where $\rho_i^0$ is the normalization constant of neutrons ($i=\text{n}$) or protons ($i=\text{p}$), $t_i$ is the diffuseness parameter, and $C_i$ is half the density radius of the neutron or proton density distribution.

\subsection{Configurational Information Entropy Method}
\label{Th:CIEM}
To determine the quantity of CIE incorporated in the fragment distributions, definitions of CIE were introduced. For a system with spatially localized clusters, when performing the CIE analysis, a set of functions $f(x)\in L^{2}(\mathbf{R})$ and their Fourier transforms $F(k)$ obey Plancherel's theorem \cite{Plantheorem}:
\begin{equation}
\int_{-\infty}^{\infty} |f(x)|^{2}\mathrm{d}x=\int_{-\infty}^{\infty} |F(k)|^{2}\mathrm{d}k,
\end{equation}
where $f(x)$ is square-integrable-bounded. The model fraction $f(k)$ is defined as:
\begin{equation}
f(k)=\frac{|F(k)|^{2}}{\int|F(k)|^{2}\mathrm{d}^{d}k},
\end{equation}
where the integration is over all $k$, $F(k)$ is defined, and $d$ is the number of spatial dimensions.

The model fraction $f(k)$ measures the relative weight of a given mode $k$. The quantity of CIE $S_\text{C}[f]$ is defined as a summation of the Shannon information entropy of $f(k)$ \cite{Shannon}:
\begin{equation}
S_\text{C}[f]=-\sum_{m=1}^{k} f_{m} \ln (f_{m}). \label{eq_Scf}
\end{equation}
Thus, the quantity of CIE has information about configurations compatible with certain constraints of a given physical system. If all the modes $k$ have the same mass, then $f_{m}=1/N$. The discrete configuration entropy reaches a maximum at $S_\text{C}= \ln N$. If there is only one mode, $S_\text{C} =$ 0.

Continuous CIE can also be defined for continuous distributions, such as the nuclear density distribution. For non-periodic functions in the interval $(a, b)$,
\begin{equation}
S_\text{C}[f]=-\int \tilde{f}(k)\ln [\tilde{f}(k)]\mathrm{d}^{d}k,
\end{equation}
where $\tilde{f}(k)=f(k)/f(k)_\text{max}$ [$f(k)_\text{max}$ is the maximum fraction]. The normalized function $\tilde{f}(k)$ guarantees that $\tilde{f}(k)\leq$ 1 for all $k$ modes, and $\tilde{f}(k) \ln \tilde{f}(k)$ denotes the CIE density.

\begin{figure}[htbp]
\includegraphics[width=\columnwidth]{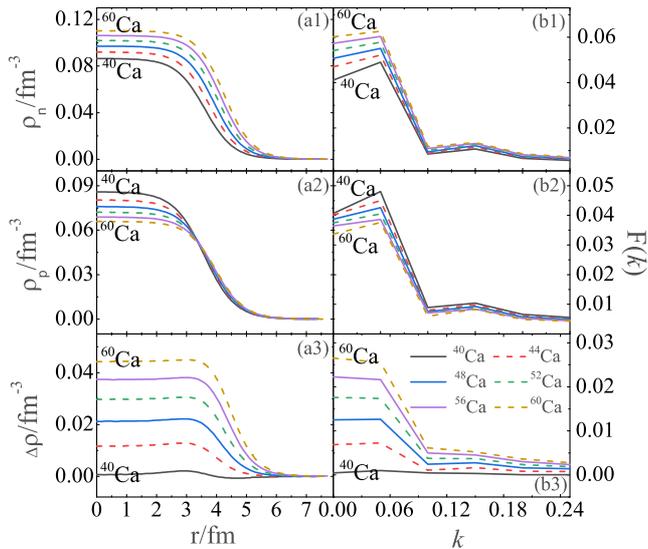}
\caption{\label{FigDensityCIE} (Color online) Fermi-type nuclear density distributions [in panels (a$j$)] and the corresponding FFT spectrum [in panels (b$j$)]. Panels (a1), (a2), and (a3) show the proton densities ($\rho_\text{p}$), neutron densities ($\rho_\text{n}$), and the nuclear density difference $\Delta\rho=\rho_\text{n}-\rho_\text{p}$ for $^{40-60}$Ca isotopes.
}
\end{figure}

\section{Results and discussion}
\label{RAD}
The 350 MeV/u $^{A_\text{p}}$Ca + $^{9}$Be reactions were calculated using the modified SAA model ($A_\text{p}$ refers to even mass numbers from 40 to 60). The cross sections of fragments with $Z$ ranging from 3 to 20 were obtained. For the sake of clarity, only part of the calculated results are shown in the figures. Figure \ref{FigDensityCIE} is a plot of the Fermi-type nuclear density distributions and their fast Fourier transformation (FFT) spectra. An obvious increase in $\rho_\text{n}$ is observed from $^{40}$Ca to $^{60}$Ca, whereas the opposite trend is observed for $\rho_\text{p}$. A two-peak structure is evident in the FFT spectra, where the second peak is lower than the first. The difference between the neutron and proton density distributions $\Delta\rho=\rho_\text{n}-\rho_\text{p}$ is also shown. For $^{40}$Ca, $\Delta\rho$ is very small, while $\Delta\rho$ increases as the neutrons  in the projectile increase. The peaks in the FFT spectra of $\rho_\text{n}$ and $\rho_\text{p}$ are not clearly shown. Based on the FFT spectra $f(k)$, the CIE of $\rho_\text{n}$, $\rho_\text{p}$, and $\Delta\rho$ can be determined using Eq. (\ref{eq_Scf}), which are denoted by $S_\text{C}^{\rho_\text{n}}[f]$, $S_\text{C}^{\rho_\text{p}}[f]$, and $S_\text{C}^{\Delta\rho}[f]$, respectively.

\begin{figure}[htbp]
\includegraphics[width=\columnwidth]{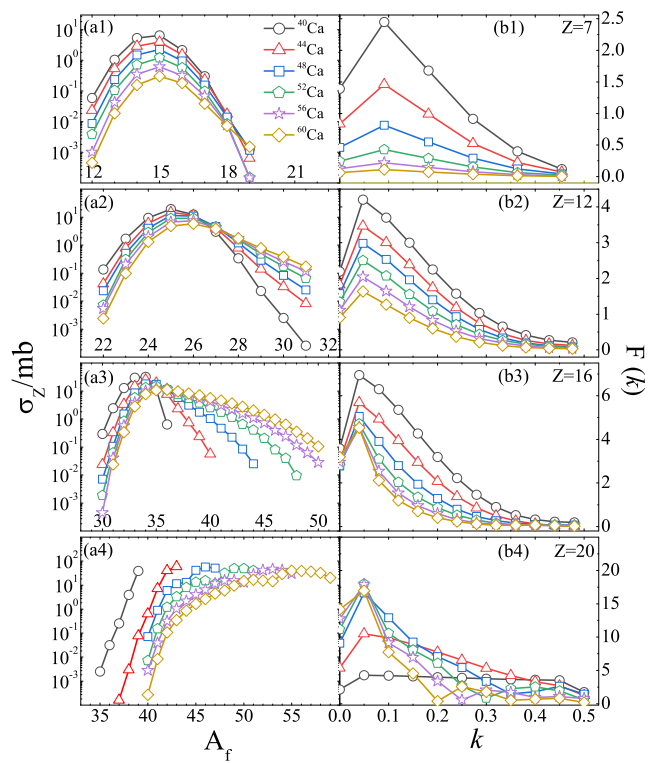}
\caption{\label{FigIsotpoic} (Color online) The isotopic cross section distributions of fragments in the 350 MeV/u $^{A_\text{p}}$Ca + $^{9}$Be reactions calculated using the modified SAA model. Panels (a$j$) correspond to $Z =$ 7, 12, 16, and 20 isotopes, respectively, and panels (b$j$) plot the corresponding FFT spectra of (ai).
}
\end{figure}

The isotopic cross section ($\sigma_Z$) distributions produced in the 350 MeV/u $^{A_\text{p}}$Ca + $^{9}$Be reactions are plotted in Fig. \ref{FigIsotpoic}. In panels (ai), from $Z_\text{fr} =$ 7 to 20, the isotopic cross-section distributions in the $^{40-60}$Ca reactions are similar for fragments with small $Z_\text{fr}$, while a shift to the neutron-rich side is observed for larger $Z_\text{fr}$. The symmetric Guassian-like shape of the isotopic distribution is altered by the enhanced cross sections of neutron-rich fragments in neutron-rich reaction systems, showing the isospin effect in fragment production induced by the increased neutron density on the surface of neutron-rich nuclei \cite{Ma09PRC}. The FFT spectra of the isotopic distributions are shown in Fig. \ref{FigIsotpoic} (b$j$).  In each FFT spectrum, only one peak is observed. The amplitudes of the FFT spectra of different isotopic distributions decrease as the projectile becomes more neutron-rich, except for $Z =$ 20. Based on the FFT spectra of $\sigma_Z$ distributions, the quantities of CIE of the isotopic distributions are determined according to Eq. (\ref{eq_Scf}), which is denoted by $S_\text{C}^{\sigma_Z}[f]$.

\begin{figure}[htbp]
\includegraphics[width=8.0cm]{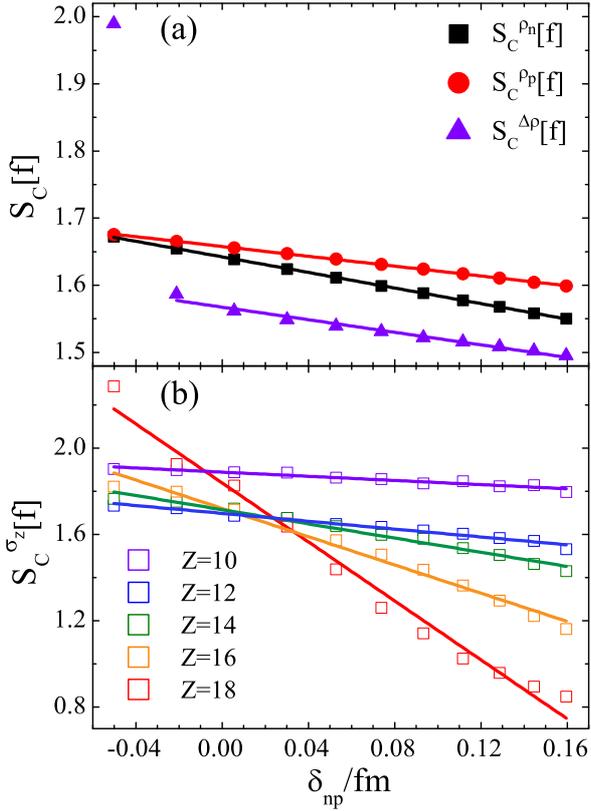}
\caption{\label{FigScDenIstp} (Color online) (a) The correlation between the CIE of $\rho_\text{n}$ (squares), $\rho_\text{p}$ (circles), $\Delta\rho$ (triangles) and neutron skin-thickness $\delta_\text{np}$ of $^{A_\text{p}}$Ca. (b) Correlation between the CIE of the isotopic cross-section distributions $S_\text{C}^{\sigma_{Z}}[f]$ and $\delta_\text{np}$ of $^{A_\text{p}}$Ca. The lines denote the linear fitting of the correlations.
}
\end{figure}

The correlation between the $S_\text{C}[f]$ of the density distribution and $\delta_\text{np}$ of the projectile nucleus is shown in Fig. \ref{FigScDenIstp}(a). Both $S_\text{C}^{\rho_\text{n}}[f]$ and $S_\text{C}^{\rho_\text{p}}[f]$ decrease linearly with an increase in $\delta_\text{np}$ from $^{40}$Ca to $^{60}$Ca. $S_\text{C}^{\Delta\rho}[f]$ also decreases linearly with increasing $\delta_\text{np}$, except for $^{40}$Ca. The correlation between the $S_\text{C}^{\sigma_Z}[f]$ of different $Z_\text{fr}$ and $\delta_\text{np}$ of the projectile nuclei are plotted in panel (b). The $S_\text{C}^{\sigma_Z}[f]$ of isotopes from $Z_\text{fr}=$ 10 to 18 are also found to decrease with the increasing $\delta_\text{np}$ of the projectile nucleus. The $S_\text{C}^{\sigma_Z}[f]$ of the fragment near the projectile nucleus was more sensitive to the change in $\delta_\text{np}$.

\begin{figure}[htbp]
\includegraphics[width=8.0cm]{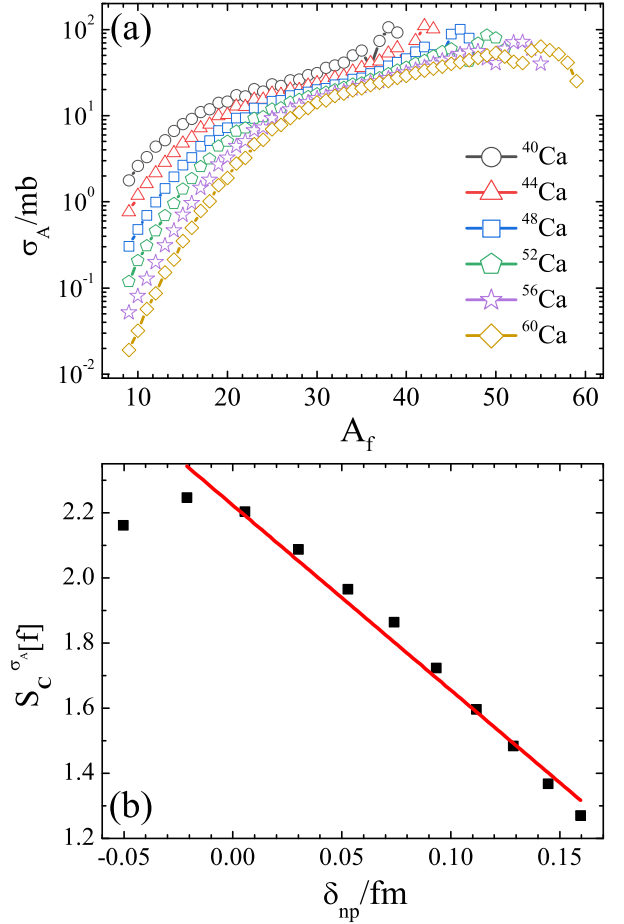}
\caption{\label{FigScMassYield} (Color online) Panel (a): The mass yield distributions of the 350 MeV/u $^{A_\text{p}}$Ca + $^{9}$Be reactions calculated using the modified SAA model. Panel (b): The correlation between $S_\text{C}^{\sigma_A}[f]$ determined from the mass yield distribution and the $\delta_\text{np}$ of $^{A_\text{p}}$Ca ($A_\text{p}$ denotes even mass numbers from 40 to 60).
}
\end{figure}

The mass yield ($\sigma_{A}$) distributions in the 350$A$ MeV $^{A_\text{p}}$Ca + $^{9}$Be reactions are shown in Fig. \ref{FigScMassYield}. In each reaction, the mass yield increases with the $A_\text{fr}$ of the fragment until it is close to the projectile nucleus. In different reactions, a very similar trend of mass distribution was observed, which decreased with the increasing mass number of the projectile nucleus. The corresponding quantities of CIE were determined from $\sigma_{A}$ distributions, which are labeled as $S_\text{C}^{\sigma_{A}}[f]$. The correlation between $S_\text{C}^{\sigma_{A}}[f]$ and $\delta_\text{np}$ for projectile nuclei is shown in Fig. \ref{FigScMassYield} (b). Except for the bend point formed at $\delta_\text{np}$ for $^{42}$Ca owing to the transition of proton-skin to neutron skin, the $S_\text{C}^{\sigma_{A}}[f] \sim \delta_\text{np}$ correlation was found to be linear for reactions of $A_\text{p} \ge$ 44.

\begin{figure}[htbp]
\includegraphics[width=8.6cm]{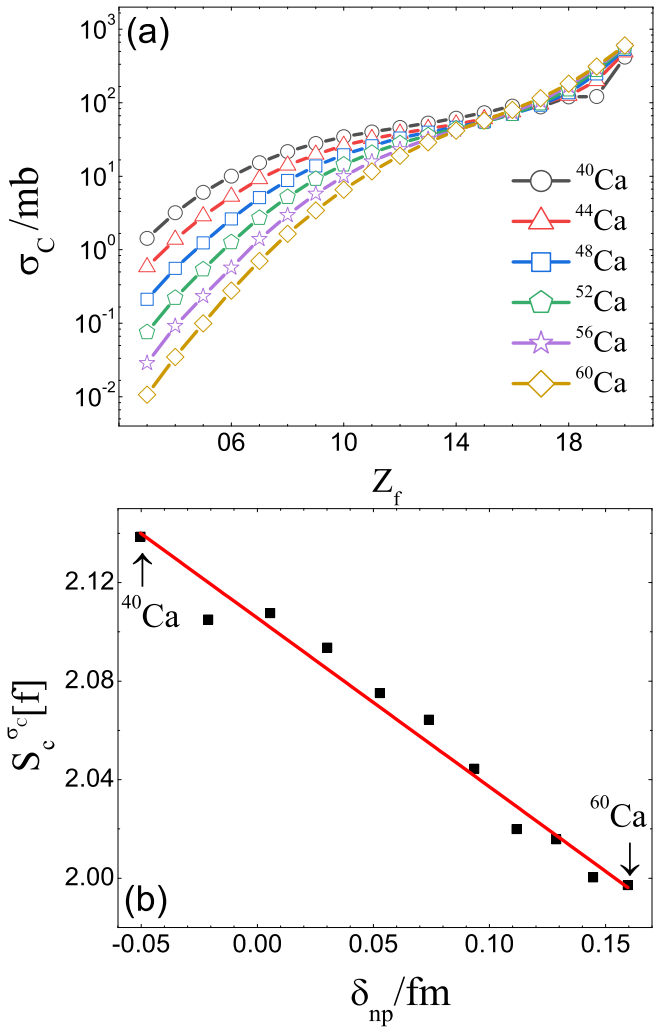}
\caption{\label{FigScChargeYield} (Color online) Similar to Fig. \ref{FigScMassYield} but for the charge cross section distributions.
}
\end{figure}

The charge cross section is defined as the summation of the isotopic cross sections $\sigma_\text{C}=\sum_{o} \sigma(A_o,Z)$. The charge cross section distributions in the 350$A$ MeV $^{A_\text{p}}$Ca + $^{9}$Be reactions are shown in Fig. \ref{FigScChargeYield}. Similar trends of $\sigma_\text{C}$ distribution trends similar to those of $\sigma_A$ were observed. The determined CIE of $\sigma_\text{C}$ distributions, labeled as $S_\text{C}^{\sigma_\text{C}}[f]$, were linearly correlated to the neutron skin thickness of the projectile nuclei.

The CIE approach transfers the experimental distributions to quantified parameters and provides information probes for determining the properties of a system. From the $S_\text{C}^{\sigma_{Z}}[f] \sim \delta_\text{np}$, $S_\text{C}^{\sigma_{A}}[f] \sim \delta_\text{np}$, and $S_\text{C}^{\sigma_\text{C}}[f] \sim \delta_\text{np}$ correlations, it was observed that the CIE determined from isotopic, mass, and charge distributions decrease with increasing neutron skin thickness, respectively, and they had good linear correlations. The determination of neutron skin thickness, in particular, nuclei near the neutron drip line, is limited by the unavailability of effective probes. The linear correlation between the CIE and neutron-skin thickness of neutron-rich nuclei provides new approaches to determine the neutron skin thickness of the projectile nucleus by measuring the fragment distributions in projectile fragmentation reactions.

\section{summary}
\label{summary}

With the vast opportunities for very asymmetric nuclei available in the new era of radioactive ion beam facilities, neutron skin thickness is one of the most important questions in nuclear physics. In this study, CIE theory is adopted to quantify the information entropy incorporated in nuclear density distributions and fragment cross-section distributions in 350 MeV/u $^{40-60}$Ca + $^{9}$Be projectile fragmentation reactions calculated using the modified SAA model. CIE quantities of nuclear density distributions ($S_\text{C}^{\rho_\text{n,p}}[f]$ and $S_\text{C}^{\Delta{\rho}}[f]$), isotopic cross-section distributions ($S_\text{C}^{\sigma_Z}[f]$), mass cross-section distributions ($S_\text{C}^{\sigma_A}[f]$), and charge cross-section distributions ($S_\text{C}^{\sigma_\text{C}}[f]$) were determined. The correlations between $S_\text{C}^{\rho_\text{p}}[f]\sim \delta_\text{np}$, $S_\text{C}^{\rho_\text{n}}[f]\sim \delta_\text{np}$, $S_\text{C}^{\Delta{\rho}}[f]\sim \delta_\text{np}$, and $S_\text{C}^{\sigma_Z}[f]\sim \delta_\text{np}$, $S_\text{C}^{\sigma_A}[f]\sim \delta_\text{np}$, and $S_\text{C}^{\sigma_C}[f]\sim \delta_\text{np}$ were also investigated. For neutron-rich calcium projectiles, obvious linear dependences of $S_\text{C}^{\rho_\text{n}}[f]$, $S_{C}^{\rho_\text{p}}[f]$, and $S_\text{C}^{\Delta{\rho}}[f]$ on $\delta_\text{np}$ were observed. The $S_\text{C}^{\sigma_Z}[f]$ of fragments with different $Z_\text{fr}$ is shown to linearly depend on the $\delta_\text{np}$ of the projectile nucleus. It is found that, if the isotopic distribution is sensitive to isospin effects in the projectiles, the extracted $S_\text{C}^{\sigma_Z}[f]$ will also be sensitive to their $\delta_\text{np}$. Good linear correlations between $S_\text{C}^{\sigma_A}[f]$, $S_\text{C}^{\sigma_\text{C}}[f]$, and $\delta_\text{np}$ of the projectile nucleus were also observed. It is suggested that, from the viewpoint of CIE, the isotopic/mass/charge distributions in the projectile fragmentation reaction may be good probes for determining the neutron-skin thickness of neutron-rich nuclei.

In this work, the simple description of the nuclear density of a projectile nucleus is difficult to deal with nuclei of magic numbers, as well as large shape distortion. Further improvements should concentrate on the inputs of nuclear densities, such as the results obtained from density functional theories, relativistic mean and field theories, to better investigate the effects of nuclear density on fragment cross-section distributions and related CIE quantities.

\section*{Author contributions}
All authors contributed to the study conception and design. Material preparation, data collection and analysis were performed by Yi-Pu Liu, Hui-Ling Wei and Chun-Wang Ma. The first draft of the manuscript was written by Yi-Pu Liu and Chun-Wang Ma and all authors commented on previous versions of the manuscript. All authors read and approved the final manuscript.


\begin{thebibliography}{99}

\bibitem{CCCS14PRC}
A. Ozawa, T. Moriguchi, T. Ohtsubo, \textup{et al}., Charge-changing cross sections of $^{30}$Ne, $^{32,33}$Na with a proton target. Phys. Rev. C {\bf 89}, 044602 (2014). \href{https://link.aps.org/doi/10.1103/PhysRevC.89.044602}{https://doi.org/10.1103/PhysRevC.89.044602}

\bibitem{CCCS19PLB}
S. Bagchi, R. Kanungo, W. Horiuchi \textup{et al}., Neutron skin and signature of the N = 14 shell gap found from measured proton radii of $^{17-22}\mathrm{N}$. Phys. Lett. B {\bf 790}, 251-256 (2019). \href{https://www.sciencedirect.com/science/article/pii/S0370269319300401}{https://doi.org/10.1016/j.physletb.2019.01.024}
	
\bibitem{Edipole15PRC}
X. Roca-Maza, X. Vi$\tilde{n}$as, M. Centelles \textup{et al}., Neutron skin thickness from the measured electric dipole polarizability in $^{68}\text{Ni}$, $^{120}\text{Sn}$, and $^{208}\text{Pb}$. Phys. Rev. C {\bf 92}, 064304 (2015). \href{https://link.aps.org/doi/10.1103/PhysRevC.92.064304}{https://doi.org/10.1103/PhysRevC.92.064304}
\bibitem{PhotonSkin15PRC}
G.F. Wei, Probing the neutron-skin thickness by photon production from reactions induced by intermediate-energy protons. Phys. Rev. C {\bf 92}, 014614 (2015). \href{https://link.aps.org/doi/10.1103/PhysRevC.92.014614}{https://doi.org/10.1103/PhysRevC.92.014614}
\bibitem{BALipi02}
B. A. Li, Probing the High Density Behavior of the Nuclear Symmetry Energy with High Energy Heavy-Ion Collisions. Phys. Rev. Lett. {\bf 88}, 192701 (2002). \href{https://link.aps.org/doi/10.1103/PhysRevLett.88.192701}{https://doi.org/10.1103/PhysRevLett.88.192701}

\bibitem{FZQ21CPL}
D.C. Zhang, H.G. Cheng, Z. Q. Feng, Hyperon Dynamics in Heavy-Ion Collisions near Threshold Energy. Chin. Phys. Lett. {\bf 38}, 092501 (2021).
\href{https://doi.org/10.1088/0256-307X/38/9/092501}{https://doi.org/10.1088/0256-307X/38/9/092501}
\bibitem{Yan19NST}
T.Z. Yan, S. Li, Impact parameter dependence of the yield ratios of light particles as a probe of neutron skin. Nucl. Sci. Tech. {\bf 30}, 43 (2019). \href{https://doi.org/10.1007/s41365-019-0572-8}{https://doi.org/10.1007/s41365-019-0572-8}
\bibitem{DecAlp17NST}
N. Wan, C. Xu, Z.Z. Ren, Exploring the sensitivity of $\alpha$-decay half-life to neutron skin thickness for nuclei around $^{208}$Pb. Nucl. Sci. Tech. {\bf 28}, 22 (2017). \href{https://doi.org/10.1007/s41365-016-0174-7}{https://doi.org/10.1007/s41365-016-0174-7}

\bibitem{HY20NST}
H. Yu, D. Q. Fang, Y.G. Ma, Investigation of the symmetry energy of nuclear matter using isospin-dependent quantum molecular dynamics. Nucl. Sci. Tech. \textbf{31}, 61 (2020).
\href{https://doi.org/10.1007/s41365-020-00766-x}{https://doi.org/10.1007/s41365-020-00766-x}
\bibitem{JL21NST}
J. Liu, C. Gao, N. Wan, C. Xu,
Basic quantities of the equation of state in isospin asymmetric nuclear matter. Nucl. Sci. Tech. \textbf{32}, 57 (2021).
\href{https://doi.org/10.1007/s41365-021-00955-2}{https://doi.org/10.1007/s41365-021-00955-2}
\bibitem{Skinisospin}
C.W. Ma, S.S. Wang, Isospin dependence of projectile fragmentation and neutron-skin thickness of neutron-rich nuclei. Chin. Phys. C {\bf 35}, 1017-1021 (2011). \href{https://doi.org/10.1088/1674-1137/35/11/007}{https://doi.org/10.1088/1674-1137/35/11/007}
\bibitem{Ma08nabr}
C.W. Ma, Y. Fu, D.Q. Fang \textup{et al}., A possible experimental observable for the determination of neutron skin thickness. Chin. Phys. B {\bf 17}, 1216-1222 (2008). \href{https://doi.org/10.1088/1674-1056/17/4/011}{https://doi.org/10.1088/1674-1056/17/4/011}
\bibitem{PRL17nremval}
T. Aumann, C.A. Bertulani, F. Schindler \textup{et al}., Peeling Off Neutron Skins from Neutron-Rich Nuclei: Constraints on the Symmetry Energy from Neutron-Removal Cross Sections. Phys. Rev. Lett. {\bf 119}, 262501 (2017). \href{https://link.aps.org/doi/10.1103/PhysRevLett.119.262501}{https://doi.org/10.1103/PhysRevLett.119.262501}

\bibitem{IYRm13PRC}
C. W. Ma, H.L. Wei, Y.G. Ma, Neutron-skin effects in isobaric yield ratios for mirror nuclei in a statistical abrasion-ablation model. Phys. Rev. C {\bf 88}, 044612 (2013). \href{https://link.aps.org/doi/10.1103/PhysRevC.88.044612}{https://doi.org/10.1103/PhysRevC.88.044612}

\bibitem{SkinIsos15PRC}
Z.T. Dai, D.Q. Fang, Y.G. Ma \textup{et al}., Effect of neutron skin thickness on projectile fragmentation. Phys. Rev. C {\bf 91}, 034618 (2015). \href{https://link.aps.org/doi/10.1103/PhysRevC.91.034618}{https://doi.org/10.1103/PhysRevC.91.034618}
\bibitem{PVS_Ren13}
J. Liu, Z. Ren, C. Xu \textup{et al}., Electroweak charge density distributions with parity-violating electron scattering. Phys. Rev. C {\bf 88}, 054321 (2013). \href{https://link.aps.org/doi/10.1103/PhysRevC.88.054321}{https://doi.org/10.1103/PhysRevC.88.054321}

\bibitem{XuJ21CPL}
J. Xu, Constraining Isovector Nuclear Interactions with Giant Dipole Resonance and Neutron Skin in Pb-208 from a Bayesian Approach. Chin. Phys. Lett. {\bf 38}, 042101 (2021).
\href{https://doi.org/10.1088/0256-307X/38/4/042101}{https://doi.org/10.1088/0256-307X/38/4/042101}

\bibitem{LRP2015USA}
Nuclear Science Advisory Committee, Reaching for the Horizon: The 2015 U.S. Long Range Plan for Nuclear Science, 2015

\bibitem{Pb208PVS}
S. Abrahamyan, Z. Ahmed, H. Albataineh \textup{et al}., Measurement of the Neutron Radius of $^{208}$Pb through Parity Violation in Electron Scattering. Phys. Rev. Lett. {\bf 108}, 112502 (2012). \href{https://link.aps.org/doi/10.1103/PhysRevLett.108.112502}{https://doi.org/10.1103/PhysRevLett.108.112502}

\bibitem{PREX2Pb}
B.T. Reed, F.J. Fattoyev, C.J. Horowitz
\textup{et al}., Implications of PREX-2 on the Equation of State of Neutron-Rich Matter. Phys. Rev. Lett. {\bf 126}, 172503 (2021). \href{https://link.aps.org/doi/10.1103/PhysRevLett.126.172503}{https://doi.org/10.1103/PhysRevLett.126.172503}

\bibitem{Shannon}
C.E. Shannon, Bell Syst. Tech. J. {\bf 27}, 379C429; 623C656  (1948).
\bibitem{CInfo16PLB}
A.E. Bernardini, R. da Rocha, Entropic information of dynamical AdS/QCD holographic models. Phys. Lett. B {\bf 762}, 107-115 (2016). \href{Uhttps://www.sciencedirect.com/science/article/pii/S0370269316305172RL}{https://doi.org/10.1016/j.physletb.2016.09.023}

\bibitem{PPNP18Ma}
C.W. Ma, Y.G. Ma, Isobaric yield ratio difference and Shannon information entropy. Prog. Part. Nucl. Phys. {\bf 99}, 120-158 (2018). \href{https://www.sciencedirect.com/science/article/pii/S0146641018300024}{https://doi.org/10.1016/j.ppnp.2018.01.002}
\bibitem{PR-Chen18}
J. Chen, D. Keane, Y.G. Ma \textup{et al.}, Antinuclei in heavy-ion collisions, Phys. Rep. {\bf 760}, 1 (2018).
\bibitem{YGMa99PRL}
Y.G. Ma, Application of Information Theory in Nuclear Liquid Gas Phase Transition. Phys. Rev. Lett. {\bf 83}, 3617 (1999). \href{https://link.aps.org/doi/10.1103/PhysRevLett.83.3617}{https://doi.org/10.1103/PhysRevLett.83.3617}
\bibitem{info15PLB}
C.W. Ma, H.L. Wei, S.S. Wang

\textup{et al.}, Isobaric yield ratio difference and Shannon information entropy.
Phys. Lett. B {\bf 742}, 19-22 (2015). \href{https://www.sciencedirect.com/science/article/pii/S0370269315000179}{https://doi.org/10.1016/j.physletb.2015.01.015}
\bibitem{info16JPG}
C.W. Ma, Y.D. Song, C.Y. Qiao
\textup{et al.}, A scaling phenomenon in the difference of Shannon information uncertainty of fragments in heavy-ion collisions.
J. Phys. G: Nucl. Part. Phys. \textbf{43}, 045102 (2016). \href{https://doi.org/10.1088/0954-3899/43/4/045102}{https://doi.org/10.1088/0954-3899/43/4/045102}
\bibitem{PPNP21Ma}
C.W. Ma, H.L. Wei, X.Q. Liu \textup{et al}., Nuclear fragments in projectile fragmentation reactions. Prog. Part. Nucl. Phys. {\bf 121}, 103991 (2021). \href{https://www.sciencedirect.com/science/article/pii/S0146641021000727}{https://doi.org/10.1016/j.ppnp.2021.103911}

\bibitem{CEDef12PLB}
M. Gleise, N. Stamatopoulos, Entropic measure for localized energy configurations: Kinks, bounces, and bubbles. Phys. Lett. B \textbf{713}, 304-307 (2012). \href{https://www.sciencedirect.com/science/article/pii/S0370269312006235}{https://doi.org/10.1016/j.physletb.2012.05.064}

\bibitem{HiggsNewbranch20NPB}
A. Alves, A.G. Dias, R. da Silva, The 7\% rule: A maximum entropy prediction on new decays of the Higgs boson. Nucl. Phys. B \textbf{959}, 115137 (2020). \href{https://www.sciencedirect.com/science/article/pii/S0550321320302236}{https://doi.org/10.1016/j.nuclphysb.2020.115137}

\bibitem{etaAds2021}
R. da Rocha, Deploying heavier meson states: Configurational entropy hybridizing AdS/QCD. Phys. Lett. B \textbf{814}, 136112 (2021). \href{https://www.sciencedirect.com/science/article/pii/S0370269321000526}{https://doi.org/10.1016/j.physletb.2021.136112}
\bibitem{QCDconftrans21}
N.R.F. Braga, O.C. Junqueira, Configuration entropy and confinement/deconfinement transiton in holographic QCD. Phys. Lett. B \textbf{814}, 136082 (2021). \href{https://www.sciencedirect.com/science/article/pii/S0370269321000228}{https://doi.org/10.1016/j.physletb.2021.136082}

\bibitem{quakniumPlasma}
N.R.F. Braga, R. da Mata, Configuration entropy for quarkonium in a finite density plasma. Phys. Rev. D \textbf{101}, 105016 (2020). \href{https://link.aps.org/doi/10.1103/PhysRevD.101.105016}{https://doi.org/10.1103/PhysRevD.101.105016}
\bibitem{timeRHIC2020}
F. Li, G. Chen, The evolution of information entropy components in relativistic heavy-ion collisions. Eur. Phys. J. A \textbf{56}, 167 (2020). \href{https://doi.org/10.1140/epja/s10050-020-00169-x}{https://doi.org/10.1140/epja/s10050-020-00169-x}

\bibitem{entropy20cpl}
J.C. Zhang, W. K. Ren, N.D. Jin, Rescaled Range Permutation Entropy: A Method for Quantifying the Dynamical Complexity of Extreme Volatility in Chaotic Time Series. Chin. Phys. Lett. \textbf{37}, 090501 (2020).
\href{https://iopscience.iop.org/article/10.1088/0256-307X/37/9/090501}{https://doi.ort/10.1088/0256-307X/37/9/090501}

\bibitem{IYRCaSkin14}
C.W. Ma, X.M. Bai, J. Yu \textup{et al}., Neutron density distributions of neutron-rich nuclei studied with the isobaric yield ratio difference. Eur. Phys. J. A {\bf 50}, 139 (2014). \href{https://doi.org/10.1140/epja/i2014-14139-1}{https://doi.org/10.1140/epja/i2014-14139-1}

\bibitem{Ma13finite}
C.W. Ma, S.S. Wang, H.L. Wei
{\it et al.}, Re-examination of Finite-Size Effects in Isobaric Yield Ratios Using a Statistical Abrasion-Ablation Model. Chin. Phys. Lett. \textbf{30}, 052101 (2013). \href{https://doi.org/10.1088/0256-307x/30/5/052501}{https://doi.org/10.1088/0256-307x/30/5/052501};

\bibitem{Ma13finite1}
C.W. Ma, H.L. Wei, Y.G. Ma, Neutron-skin effects in isobaric yield ratios for mirror nuclei in a statistical abrasion-ablation model. Phys. Rev. C \textbf{88}, 044612 (2013). \href{https://link.aps.org/doi/10.1103/PhysRevC.88.044612}{https://doi.org/10.1103/PhysRevC.88.044612}
\bibitem{FANG00}
D.Q. Fang, W.Q. Shen, J. Feng \textup{et al}., Isospin effect of fragmentation reactions induced by intermediate energy heavy ions and its disappearance. Phys. Rev. C {\bf 61}, 044610 (2000). \href{https://link.aps.org/doi/10.1103/PhysRevC.61.044610}{https://doi.org/10.1103/PhysRevC.61.044610}
\bibitem{Ma09PRC}
C.W. Ma, J.Y. Wang, H.L. Wei \textup{et al}., Isospin dependence of projectile-like fragment production at intermediate energies. Phys. Rev. C {\bf 79}, 034606 (2009). \href{https://link.aps.org/doi/10.1103/PhysRevC.79.034606}{https://doi.org/10.1103/PhysRevC.79.034606}
\bibitem{Brohm94}
T. Brohm, K.-H. Schmidt, Statistical abrasion of nucleons from realistic nuclear-matter distributions. Nucl. Phys. A {\bf 569}, 821-832 (1994). \href{https://www.sciencedirect.com/science/article/pii/0375947494903867}{https://doi.org/10.1016/0375-9474(94)90386-7}
\bibitem{EIS54}
Y. Eisenberg, Interaction of Heavy Primary Cosmic Rays in Lead. Phys. Rev. {\bf 96} 1378 (1954). \href{https://link.aps.org/doi/10.1103/PhysRev.96.1378}{https://doi.org/10.1103/PhysRev.96.1378}
\bibitem{GAIM91}
J.J. Gaimard, K.-H. Schmidt, A reexamination of the abrasion-ablation model for the description of the nuclear fragmentation reaction. Nucl. Phys. A {\bf 531}, 709-745 (1991). \href{https://www.sciencedirect.com/science/article/pii/037594749190748U}{https://doi.org/10.1016/0375-9474(91)90748-U}
\bibitem{Cai98}
X. Cai, J. Feng, W. Shen \textup{et al}., In-medium nucleon-nucleon cross section and its effect on total nuclear reaction cross section. Phys. Rev. C {\bf 58}, 572-575 (1998). \href{https://link.aps.org/doi/10.1103/PhysRevC.58.572}{https://doi.org/10.1103/PhysRevC.58.572}

\bibitem{Plantheorem}
W. Rudin,  Real and Complex Analysis, McGrawCHill, 1987, Singapore


\end{thebibliography}
\end{document}